\documentclass[twocolumn,showpacs,preprintnumbers,amsmath,amssymb]{revtex4}

\usepackage{graphicx}
\usepackage{dcolumn}
\usepackage{bm}
\usepackage{wrapfig}

\begin{document}

\preprint{APS/123-QED}

\title{
Nodal superconductivity and non-Fermi-liquid behavior in Ce$_{2}$PdIn$_{8}$ \\
studied by $^{115}$In nuclear quadrupole resonance
}

\author{
H. Fukazawa~\footnote{hideto@nmr.s.chiba-u.ac.jp}, R. Nagashima, S. Shimatani, and Y. Kohori
}
\address{
Department of Physics, Graduate School of Science, Chiba University, Chiba 263-8522, Japan
}

\author{
D. Kaczorowski
}
\address{
Institute of Low Temperature and Structure Research, Polish Academy of Sciences, P. O. Box 1410, 50-950 Wroc{\l}aw, Poland
}
%
%
%
%
\begin{abstract}
Nuclear quadrupole resonance measurements were performed on the heavy fermion superconductor Ce$_{2}$PdIn$_{8}$. Above the Kondo coherence temperature $T_{\rm coh}\simeq 30$~K, the spin-lattice relaxation rate $1/T_{1}$ is temperature independent, whereas at lower temperatures, down to the onset of superconductivity at $T_{\rm c} = 0.64$~K, it is nearly proportional to $T^{1/2}$. Below $T_{\rm c}$, $1/T_{1}$ shows no coherence peak and decreases as $T^{3}$ down to 75~mK. All these findings
indicate that Ce$_{2}$PdIn$_{8}$ is close to the antiferromagnetic quantum critical point, and the superconducting state has an unconventional character with line nodes in the superconducting gap.
\end{abstract}
\maketitle

\section{Introduction}

The interplay between superconductivity and magnetism is one of the central issues in strongly correlated electron systems (SCES)~\cite{Sac1}. An excellent playground for the systematic study of the mutual relationship between the two phenomena is a series of Ce-based heavy-fermion compounds with the formula Ce$_{n}M$In$_{3n+2}$ ($n = 1,2,\infty,\, M=$~Co, Rh, Ir, Pd)~\cite{Math,Pet1,Heg1,Pet2,Che1,Kac1,Koh1,Fuk4,Zhe1}. In these intermetallics, the magnetic correlations arise from the Ruderman-Kittel-Kasuya-Yosida interaction and the Kondo interaction, which are both driven by the hybridization of localized $4f$ electrons and itinerant conduction electrons. Unconventional superconductivity has been discovered, e.g., in pressurized CeIn$_{3}$~\cite{Math} and CeRhIn$_{5}$,~\cite{Heg1} as well as in CeCoIn$_{5}$ under ambient pressure conditions.~\cite{Pet1} In each case the superconductivity emerges near a magnetic instability at $T$ = 0, i.e., the quantum critical point (QCP). Other Ce$_{n}M$In$_{3n+2}$ phases are also good candidates for the investigation of mutual relationship among dimensionality, quantum criticality, and superconductivity, because their crystal structures can be viewed as a regular stacking of the three-dimensional (3D) CeIn$_{3}$ and quasi-two-dimensional (quasi-2D) Ce$M$In$_{5}$ units (see Fig.~1).

\begin{figure}
\includegraphics[width=8cm]{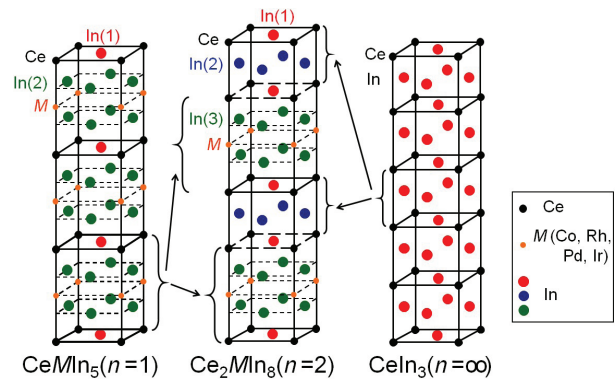}
\caption{
(Color online) Crystal structures of the Ce$_{n}M$In$_{3n+2}$ ($M=$~Co, Rh, Ir, Pd) compounds.
}
\label{f1}
\end{figure}

Recently, the compound Ce$_{2}$PdIn$_{8}$ was reported to exhibit normal- and superconducting (SC)-state properties very similar to those of CeCoIn$_{5}$.~\cite{Kac1,Kac2,Kac3,Mat1,Gni1,Tra1,Don1,Tok1} This compound is a heavy-fermion superconductor with $T_{\rm c}\simeq$~0.7~K and the normal-state electronic specific heat coefficient of approximately 1.5~J/K$^{2}$ mol-fu.~\cite{Kac1}. Above $T_{\rm c}$, the electrical transport shows distinct non-Fermi-liquid (NFL) behavior, i.e., linear-in-$T$ electrical resistivity, highly anomalous $T$- and $H$-dependent magnetoresistance, and Hall, Nernst, and Seebeck effects.~\cite{Kac1,Kac2,Kac3,Mat1,Gni1} In parallel, the electronic specific heat ratio $C/T$ shows a logarithmic temperature dependence.~\cite{Kac1,Kac2,Kac3,Tok1} Moreover, it has been established that the hydrostatic pressure or magnetic field can tune $T_{\rm c}$ towards $T$ = 0, and clear evidence for the magnetic field-induced QCP near the upper critical field $H_{\rm c2}\simeq$~2.3 T has been found.~\cite{Tra1,Don1} As for the superconductivity, it has been shown that the most probable candidate for the SC gap symmetry is the $d$-wave.~\cite{Don1,Tok1} Thermal conductivity measurements in applied magnetic fields have revealed the presence of residual density of states (DOS) at the Fermi level.~\cite{Don1} Furthermore, recent penetration depth measurements using a tunnel diode resonator have also indicated a nodal-line structure within the SC energy gap.~\cite{Has1} Both $T_{\rm c}$ and $H_{\rm c2}$ are considerably lower than those derived for CeCoIn$_{5}$,~\cite{Heg1,Koh1} and this difference is probably a direct consequence of the fact that the crystal lattice of Ce$_{2}$PdIn$_{8}$ has a much less 2D character than that of CeCoIn$_{5}$.

\begin{figure*}
\includegraphics[width=12cm]{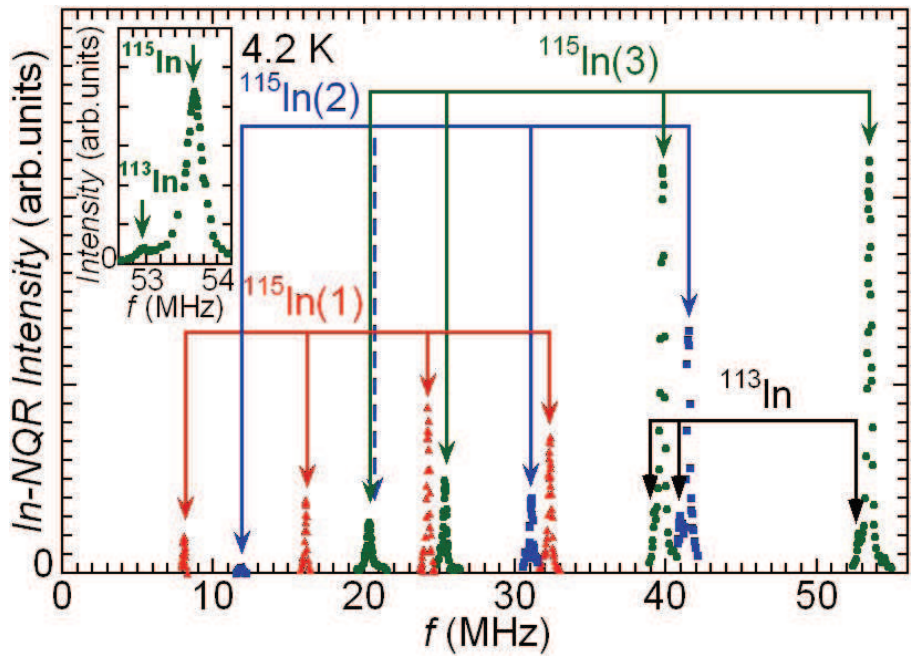}
\caption{
(Color online) In-NQR spectrum of Ce$_{2}$PdIn$_{8}$.~\cite{Not1}
Inset: 4$\nu_{Q}$ lines of In(3) site for $^{113}$In and $^{115}$In nuclei.
}
\label{f2}
\end{figure*}

The fascinating phenomena in Ce$_{2}$PdIn$_{8}$ have been revealed  by bulk property measurements. To the best of our knowledge, no microscopic or site-selective studies have been performed so far, which would contribute to a better characterization of the magnetic properties in the normal state and the unconventional SC behavior at the lowest temperatures. In this paper, we report on our $^{115}$In nuclear quadrupole resonance (NQR) measurements of Ce$_{2}$PdIn$_{8}$. The NQR data corroborate the aforepostulated quantum critical and nodal SC features in this compound.

\section{Experiments}

A polycrystalline sample of Ce$_{2}$PdIn$_{8}$ was synthesized by the arc melting method, as described in Ref.~\cite{Kac3}. The quality of the product was checked by x-ray powder diffraction and energy-dispersive x-ray analysis, as well as by measurements of the bulk properties. The sample was found to be a single-phase with a fairly sharp SC transition at $T_{\rm c}\simeq$~ 0.65~K.~\cite{Kac1} For NQR measurements the ingot was powdered in order to reduce the heating-up effect at low temperatures and improve the signal intensity. The $^{115}$In NQR studies were performed in the frequency range of 7$-$60~MHz using a phase-coherent pulsed NQR spectrometer. Measurements were carried out above 1.5~K using a $^{4}$He cryostat and below 1.5~K with a $^{3}$He-$^{4}$He dilution refrigerator. The spin-lattice relaxation time $T_{1}$ was obtained from the recovery of the nuclear magnetization after a saturation pulse. $T_{\rm c}$ was estimated to be 0.64~K through ac susceptibility by measuring the temperature dependence of the characteristic frequency of the NQR sample coil $f_{\rm coil}$ (see the inset in Fig.~\ref{f5}).

\section{Results and Discussion}

Figure~\ref{f2} shows the NQR spectrum of Ce$_{2}$PdIn$_{8}$ taken at 4.2~K. As can be inferred from Fig.~\ref{f1}, there are three inequivalent crystallographic positions for indium atoms in the unit cell of Ce$_{2}$PdIn$_{8}$: the most symmetric site, In(1) ($4mm$); the less symmetric site, In(2) ($mmm.$); and the least symmetric site, In(3) ($2mm.$). For the In nuclei ($I$ = 9/2), the electric quadrupole Hamiltonian $\mathcal{H}_{Q}$ is given by
\begin{equation}
\mathcal{H}_{Q} = \frac{e^{2}qQ}{4I(2I-1)}\Bigl\{ 3I_{z}^{2}-I(I+1)+\frac{\eta}{2}(I_{+}^{2}+I_{-}^{2}) \Bigl\},
\end{equation}
where $eq$, $eQ$ and $\eta$ represent the electric field gradient (EFG), the nuclear quadrupole moment, and the asymmetry parameter of the EFG, respectively. By diagonalizing $\mathcal{H}_{Q}$ and considering the three inequivalent In sites, one can assign all the observed features in the NQR spectrum. For each In site, four resonance lines occur with increasing frequency: $1\nu_{Q}$, $2\nu_{Q}$, $3\nu_{Q}$ and $4\nu_{Q}$. The $1\nu_{Q}$ line corresponds to the transition between $|I_{z}=\pm 1/2\rangle$ and $|I_{z}=\pm 3/2\rangle$, the $2\nu_{Q}$ line is due to the transition between $|I_{z}=\pm 3/2\rangle$ and $|I_{z}=\pm 5/2\rangle$, etc. Note that $1\nu_{Q}$ differs from the resonance frequency $\nu_{0}\equiv\frac{3e^{2}qQ}{2I(2I-1)}$ except for the case of $\eta = 0$. The obtained NQR frequency $\nu_{\rm 0}$ and the parameter $\eta$ are 8.08~MHz and 0.00(1), respectively, for In(1); 10.37~MHz and 0.12(1) for In(2); and 13.37~MHz and 0.27(1) for In(3). At around 20.4~MHz, the 2$\nu_{\rm Q}$ line for In(2) and the 1$\nu_{\rm Q}$ line for In(3) accidentally coincide with each other.

As is apparent from Fig.~\ref{f2}, no NQR signals were observed other than those arising from the $^{113}$In and $^{115}$In nuclei in Ce$_{2}$PdIn$_{8}$.~\cite{Not1} This finding confirms the quality of the specimen, which was free of any clear parasitic phases that were found in Ce$_{2}$CoIn$_{8}$ and Ce$_{2}$RhIn$_{8}$.~\cite{Fuk4} The line-width is about 1.8 times wider than the corresponding line-widths of single-crystalline CeCoIn$_{5}$~\cite{Koh1} and Ce$_{2}$CoIn$_{8}$~\cite{Fuk4}. This line broadening may signal some structural disorder in the sample studied and/or may arise due to averaging EFG in the powder sample. With decreasing temperature, the resonance frequency increases and then becomes nearly constant below approximately 10~K. The width of particular lines does not change with decreasing temperature, since the compound remains paramagnetic down to the lowest temperature investigated.

\begin{figure}
\includegraphics[width=8cm]{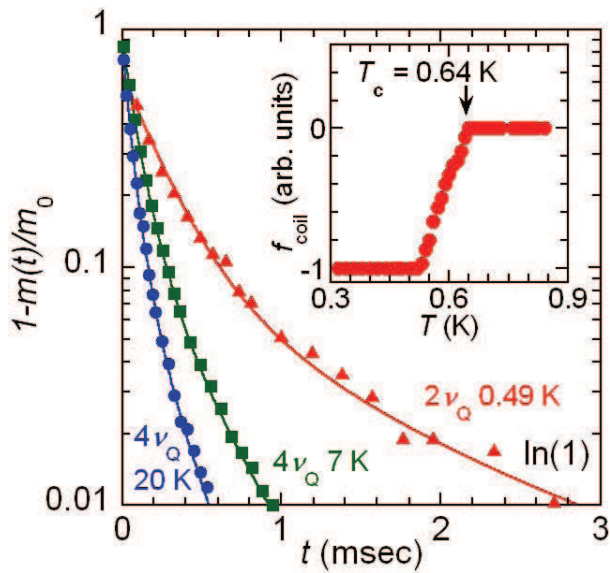}
\caption{
(Color online) Nuclear magnetization recovery curves of Ce$_{2}$PdIn$_{8}$ at 0.45, 7, and 20~K. Solid lines denote the fitting curves discussed in the text. Inset: Temperature dependence of the characteristic frequency of the NQR sample coil $f_{\rm coil}$.
}
\label{f5}
\end{figure}

The spin-lattice relaxation time $T_{1}$ was derived over the range of 0.075$-$150~K at the frequency of the spectral center at each In site. The nuclear magnetization recovery curves were analyzed in terms of the quadruple exponential function appropriate for the nuclear spin $I=9/2$ of the $^{115}$In nuclei~\cite{Mac1}. For example, for the 4$\nu_{\rm Q}$ line with $\eta = 0$ the following formula was applied,
\begin{eqnarray}
1-\frac{m(t)}{m_{0}} = 0.1212\exp\left( -\frac{3t}{T_{1}}\right)  + 0.5594\exp\left( -\frac{10t}{T_{1}}\right)\nonumber \\
+0.297\exp\left( -\frac{21t}{T_{1}}\right)  + 0.0224\exp\left( -\frac{36t}{T_{1}}\right),
\end{eqnarray}
where $m(t)$ and $m_{0}$ denote the nuclear magnetization after a time $t$ from the NMR saturation pulse and the thermal equilibrium magnetization, respectively. For In sites with a finite $\eta$, the modification of the recovery curve was adopted from Ref.~\cite{Chep}. The above theoretical formula accounts for magnetic relaxation only, and the recovery curve becomes more complicated if quadrupolar relaxation is involved.~\cite{Mac1} As is apparent from Fig.~\ref{f5}, all the experimental data, even those taken at the lowest temperatures, can be well approximated by this function using a single-component $T_{1}$. This means that indeed the magnetic relaxation is a primary contribution to the behavior of the compound studied.

\begin{figure}
\includegraphics[width=8cm]{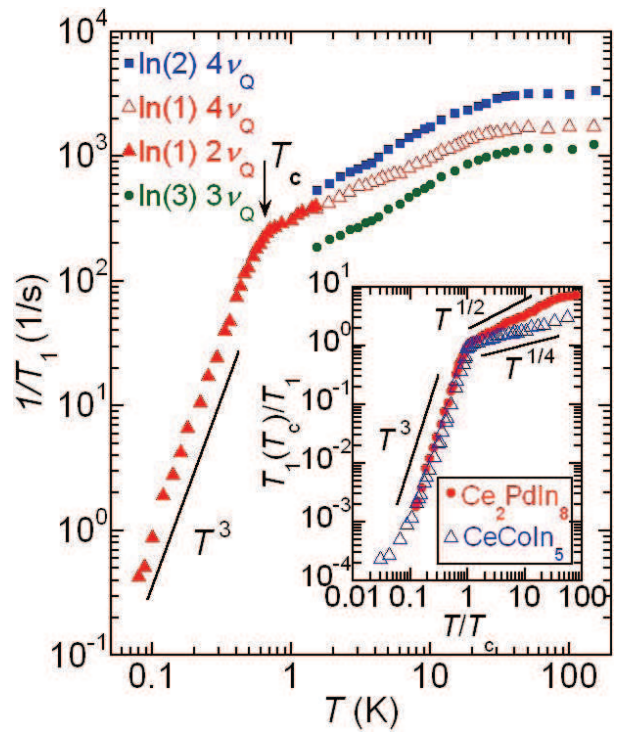}
\caption{
(Color online) Temperature dependence of the spin-lattice relaxation rate $1/T_{1}$ at the three inequivalent In sites of Ce$_{2}$PdIn$_{8}$. Inset: Temperature dependencies of $1/T_{1}$ at the In(1) site of Ce$_{2}$PdIn$_{8}$ and CeCoIn$_{5}$, normalized at the respective $T_{\rm c}$. The $1/T_{1}$ data on CeCoIn$_{5}$ were taken from Ref.~\cite{Koh1}.
}
\label{f3}
\end{figure}

The so-obtained temperature dependencies of the spin-lattice relaxation rate $1/T_{1}$ in Ce$_{2}$PdIn$_{8}$ are displayed in Fig.~\ref{f3}. For the In(1) site, $1/T_{1}$ was evaluated from the $2\nu_{\rm Q}$ and $4\nu_{\rm Q}$ lines, and the results were found to perfectly overlap at 1.5~K, thus proving the reliability of the determination of $1/T_{1}$. As shown in Fig.~\ref{f3}, the relaxation rate $1/T_{1}$ is largest at In(2), and smallest at In(3). This tendency is the same as that of isostructural Ce$_{2}$CoIn$_{8}$ and Ce$_{2}$RhIn$_{8}$.~\cite{Fuk4}

From the bulk property data, Ce$_{2}$PdIn$_{8}$ has been characterized as a Kondo lattice with the characteristic temperature ($T_{\rm coh}\simeq$~30~K) that signals the crossover from the high-temperature incoherent to low-temperature coherent Kondo state~\cite{Kac1,Kac2,Kac3}. In line with that scenario, the relaxation rate $1/T_{1}$ is nearly temperature independent above $T_{\rm coh}$, as expected for fairly well localized $4f$ electrons of trivalent Ce ions. Below $T_{\rm coh}$ and down to 0.7 K $(\simeq T_{\rm c})$, $1/T_{1}$ varies approximately as $T^{1/2}$. This temperature dependence is distinctly different from the Korringa relation $T_{1}T$ = const, appropriate for Fermi liquids. In turn, it is quantitatively similar to that of CeIrIn$_{5}$, which exhibits well-established NFL behavior in the normal state.~\cite{Koh1,Zhe1} Thus, the $1/T_{1}$ data on Ce$_{2}$PdIn$_{8}$ are fully consistent with the NFL features observed in the bulk properties of the compound.~\cite{Kac1,Kac2,Kac3,Mat1,Gni1,Tra1,Don1,Tok1}

At the onset of the SC state, $1/T_{1}$ decreases rapidly, without the Hebel-Slichter coherence peak, and then varies nearly proportionally to $T^{3}$ down to the lowest temperatures studied. This behavior is quite similar to that found for CeCoIn$_{5}$ (see the inset in Fig.~\ref{f3}) and CeIrIn$_{5}$,~\cite{Koh1,Zhe1} and it hints at a nodal character of the SC energy gap. It is worth noting that down to the lowest measurement temperature of 75~mK ($\approx T_{\rm c}/9$), no saturation in $1/T_{1}(T)$ is seen, which would be caused by some admixture of paramagnetic impurities. 

The $T^{3}$ dependence of $1/T_{1}$ in the SC state, in conjunction with the absence of the Hebel-Slichter peak below $T_{\rm c}$, indicates that the SC gap in Ce$_{2}$PdIn$_{8}$ has a nodal character. In general, the nuclear spin-lattice relaxation rate in the SC state can be expressed as
\begin{eqnarray}
1/T_{1} \propto \int_{0}^{\infty}\{ N_{\rm S}(E,\Delta(T))^{2}+M_{\rm S}(E,\Delta(T))^{2}\} \nonumber\\
\times f(E)\{ 1-f(E)\} dE,
\end{eqnarray}
where $N_{\rm S}(E,\Delta(T))$ is the DOS, $M_{\rm S}(E,\Delta(T))$ is the anomalous DOS arising from the coherence effect of the Cooper pairs, and $f(E)$ stands for the Fermi distribution function. The energy gap $\Delta(T)$ in this equation can be anisotropic, and thus more precisely it should be written as $\Delta(T,\theta,\varphi)$. Assuming that the SC gap has a simple horizontal line node in $q$-space, i.e., $\Delta(T,\theta,\varphi)=\Delta(T)\cos\theta$, and taking into account that the integral of $M_{\rm S}(E,\Delta(T))$ effectively vanishes due to the alternating sign change in $\Delta(T,\theta,\varphi)$, one can properly reproduce the experimental $1/T_{1}(T)$ data on Ce$_{2}$PdIn$_{8}$ in the entire SC range studied (see the inset in Fig.~\ref{f3}). On the contrary, no satisfactory description of $1/T_{1}(T)$ could be obtained with an isotropic SC gap, even if the anomalous DOS term is omitted. The calculations yielded $2\Delta(0)/k_{\rm B}T_{\rm c} = 5.6$, which is very close to the $2\Delta(0)/k_{\rm B}T_{\rm c} = 6.4$ derived for CeCoIn$_{5}$ by means of the analogous analysis of the NQR data.~\cite{Koh1} The similarity in the SC gap amplitudes in the two compounds provides more support for the hypothesis on the $d$-wave symmetry of the SC order parameter in Ce$_{2}$PdIn$_{8}$, formulated before on the basis of the bulk property data.~\cite{Don1,Tok1} However, in order to reach a definitive conclusion on this issue, further experimental efforts are necessary, among them NMR Knight shift measurements performed on single-crystalline samples.

\section{Summary}

An NQR study on the heavy-fermion superconductor Ce$_{2}$PdIn$_{8}$ has revealed distinct NFL behavior of the nuclear spin-lattice relaxation above the SC critical temperature $T_{\rm c} = 0.64$~K. The temperature dependence $1/T_{1} \propto T^{1/2}$, observed up to $T_{\rm coh}\sim 30$~K, indicates that the compound is located close to the antiferromagnetic QCP instability. In the SC state, $1/T_{1}$ shows no coherence peak and it is proportional to $T^{3}$ at least down to 75~mK. These features manifest unconventional superconductivity with line nodes in the SC gap, probably of a character similar to that in CeCoIn$_{5}$. Further NMR/NQR studies in applied magnetic fields and/or under high pressure, performed on single crystals, would be highly beneficial for microscopic characterization of the properties of Ce$_{2}$PdIn$_{8}$ near the field- or pressure-induced QCP.

\vspace{5mm}
\begin{center}

{\bf ACKNOWLEDGMENTS}

\end{center}

\vspace{5mm}

The authors thank S. Ohara for fruitful discussion. This work was supported by Grants-in-Aid for Scientific Research (Nos. 21540351 and 22684016) from MEXT and JSPS, Innovative Areas ``Heavy Electrons" (No. 21102505) from MEXT, and the Global COE and AGSST financial support program of Chiba University. The work in Poland was supported by the National Science Centre (Poland) under Research Grant No. 2011/01/B/ST3/04482.

\newpage

\begin{thebibliography}{99}

\bibitem{Sac1}
S. Sachdev and B. Keimer, Phys. Today {\bf 64}, 29 (2011).

\bibitem{Math}
N. D. Mathur, F. M. Grosche, S. R. Julian, I. R. Walker, D. M. Freye,
R. K. W. Haselwimmer, and G. G. Lonzarich, Nature {\bf 394}, 39 (1998).

\bibitem{Heg1}
H. Hegger, C. Petrovic, E. G. Moshopoulou, M. F. Hundley, J. L. Sarrao, Z. Fisk, and J. D. Thompson,
Phys. Rev. Lett. {\bf 84}, 4986 (2000).

\bibitem{Pet1}
C. Petrovic, P. G. Pagliuso, M. F. Hundley, R. Movshovich, J. L. Sarrao,
J. D. Thompson, Z. Fisk, and P. Monthoux, J. Phys.: Condens. Matter {\bf 13}, L337 (2001).

\bibitem{Pet2}
C. Petrovic, R. Movshovich, M. Jaime, P. G. Pagliuso, M. F. Hundley, J. L. Sarrao, Z. Fisk, and J. D. Thompson,
Europhys. Lett. {\bf 53}, 354 (2001).

\bibitem{Che1}
G. Chen, S. Ohara, M. Hedo, Y. Uwatoko, K. Saito, M. Sorai, and I. Sakamoto, J. Phys. Soc. Jpn. {\bf 71}, 2836 (2002).

\bibitem{Kac1}
D. Kaczorowski, A. P. Pikul, D. Gnida, and V. H. Tran, Phys. Rev. Lett. {\bf 103}, 027003 (2009):{\it ibid} {\bf 104}, 059702 (2010).

\bibitem{Koh1}
Y. Kohori, Y. Yamato, Y. Iwamoto, T. Kohara, E. D. Bauer, M. B. Maple, and J. L. Sarrao: Phys. Rev. B {\bf 64} (2001) 134526.

\bibitem{Zhe1}
G.-Q. Zheng, K. Tanabe, T. Mito, S. Kawasaki, Y. Kitaoka, D. Aoki, Y. Haga, and Y. Onuki, Phys. Rev. Lett. {\bf 86}, 4664 (2001).

\bibitem{Fuk4}
H. Fukazawa, T. Okazaki, K. Hirayama, Y. Kohori, G. Chen, S. Ohara, I. Sakamoto, and T. Matusmoto,
J. Phys. Soc. Jpn. {\bf 76} (2007) 124703.

\bibitem{Kac2}
D. Kaczorowski, A. P. Pikul, B. Belan, L. Sojka, and Ya. Kalychak, Physica B {\bf404}, 2975 (2009).

\bibitem{Kac3}
D. Kaczorowski, D. Gnida, A. P. Pikul, and V. H. Tran, Solid State Commun. {\bf150}, 411 (2010).

\bibitem{Mat1}
M. Matusiak, D. Gnida, and D. Kaczorowski, Phys. Rev. B {\bf 84}, 115110 (2011).

\bibitem{Gni1}
D. Gnida, M. Matusiak, and D. Kaczorowski, Phys. Rev. B {\bf 85}, 060508(R) (2012).

\bibitem{Tok1}
Y. Tokiwa, P. Gegenwart, D. Gnida, and D. Kaczorowski, Phys. Rev. B {\bf 84}, 140507(R) (2011).

\bibitem{Tra1}
V. H. Tran, D. Kaczorowski, R. T. Khan, and E. Bauer, Phys. Rev. B {\bf 83}, 064504 (2011).

\bibitem{Don1}
J. K. Dong, H. Zhang, X. Qiu, B. Y. Pan, Y. F. Dai, T. Y. Guan, S. Y. Zhou, D. Gnida, D. Kaczorowski, and S.Y. Li,
Phys. Rev. X {\bf 1}, 011010 (2011).

\bibitem{Has1}
K. Hashimoto {\it et al.} (unpublished).

\bibitem{Not1} Because the natural abundance of $^{113}$In is 4.3\% and that of $^{115}$In is 95.7\%,
the $^{113}$In line intensities are quite low (see the inset in Fig.~\ref{f2}). 
The lines due to $^{113}$In overlap with those of $^{115}$In because the ratio of quadrupole moments is $^{113}Q/^{115}Q = 0.82/0.83$. 
However, they can be identified because of this ratio and the natural-abundance ratio. 

\bibitem{Mac1}
D. E. MacLaughlin, J. D. Williamson, and J. Butterworth, Phys. Rev. B {\bf 4}, 60 (1971).

\bibitem{Chep}
J. Chepin and J. H. Ross, J. Phys.: Condens. Matter {\bf 3}, 8103 (1991).




\end{thebibliography}
\end{document}